\documentclass[11pt]{article}
\usepackage{amscd}
\usepackage{amssymb}
\usepackage{amsfonts}
\usepackage{amsmath}
\usepackage{mathrsfs}
\begin{document}

\setcounter{page}{1}
\vfil

\pagestyle{plain}

\begin{center}
{\LARGE {\bf Scale symmetry in classical and quantum mechanics}}

\bigskip
\bigskip
\bigskip

{\large E. Gozzi} and {\large D. Mauro}

             Department of Theoretical Physics, 
	     University of Trieste, \\
	     Strada Costiera 11, Miramare-Grignano, 34014 Trieste, Italy,\\
	     and INFN, Trieste, Italy\\
	     e-mail: {\it gozzi@ts.infn.it} and  {\it mauro@ts.infn.it}

\end{center}

\bigskip
\bigskip
\begin{abstract}
\noindent In this paper we address again the issue of the scale anomaly in quantum mechanical models with inverse square potential. In particular we examine the interplay between the classical and quantum aspects of the system using in both cases an operatorial approach. 
\end{abstract}

\section{Introduction}

Scale symmetry has always played an important role in physics, especially in field theories where it is broken by the renormalization effects. Lately this symmetry has been studied also in quantum mechanical models with a finite number of degrees of freedom. The two mechanical models which present scale invariance at the classical level are the inverse square potential \cite{alfaro}-\cite{gupta}  and the Dirac delta interaction in two dimensions \cite{jackiw}. In this paper we will concentrate on the first kind of potential. Since the first paper in 1950 \cite{furlan} a lot of work has been devoted to this potential
\cite{cambl}.

It has been proved \cite{cambl} that in the strong coupling regime the classical scale invariance of the inverse square potential is broken by the quantization procedure. This is due to the fact that the Hamiltonian is not a self-adjoint operator \cite{esteve}. Another approach has been based on the fact that the spectrum of the inverse square Hamiltonian is not bounded from below and a sort of ``regularization" procedure is needed \cite{gupta} to get a stable ground state. After this regularization is implemented, the scale invariance gets broken and a quantum anomaly arises. Usually anomalies arise only in field theories and the fact that in this case they take place in a system with a finite number of degrees of freedom is something worth investigating further. The analysis we will present here is centered on the classical aspects of the model. The reason for this investigation is because another potential, the Dirac delta one, which was thought to develop a scale anomaly at the quantum level, was proved to have problems already in the classical regime \cite{cabo}. Basically it was shown that, according to which representation of the Dirac delta one uses, the scale symmetry could already be broken at the classical level. The lesson we have learned from this case is that, before going to the quantum regime, it is better to study further the system at the classical level which is what we shall do in this paper for the inverse square potential. 

The formalism for classical mechanics (CM) that we shall use in this analysis is not so well-known in the particle physics community and so we will briefly review it in the next section. This method goes under the name of Koopman-von Neumann (KvN) operatorial approach to CM \cite{koopman}.
Despite its operatorial nature, the KvN formalism is still classical mechanics and so the scale symmetry in the inverse square potential must be unbroken. We thought of using the KvN method because, being operatorial in nature, it makes CM very similar to quantum mechanics (QM) and so it may help in throwing light on the mechanism that makes the anomaly appear in QM and disappear in CM. In a previous paper \cite{chir} we have studied in the KvN formalism the chiral symmetry, which at the quantum level is anomalous. We showed that, due to an enlargement in the number of variables appearing in the KvN with respect to the quantum mechanical case, the symmetry is maintained. In fact at the classical level a sort of ``cancellation" of the anomaly takes place between the different sets of KvN variables. We will show in this Letter that something similar, even if not exactly the same, happens for the scale anomaly. 

\section{KvN formulation of CM}

In order to make this Letter self-contained, let us start by briefly reviewing some aspects of the KvN formulation. In classical statistical mechanics the time evolution of the classical probability densities in phase space $\rho(\vec{r},\vec{p},t)$ is given by the so called Liouville equation:
\begin{equation}
\displaystyle i\frac{\partial}{\partial t} \rho \left(\vec{r},\vec{p},t \right)= 
\widehat{\cal H} \rho \left( \vec{r},\vec{p},t\right) \label{liouveq}
\end{equation}
where the Liouvillian $\widehat{\cal H}$ is:
\begin{equation}
\widehat{\cal H}=
-i\vec{\partial}_{p} H(\vec{r},\vec{p}\,) \cdot \vec{\partial}_r+
i\vec{\partial}_r H(\vec{r},\vec{p}\,) \cdot \vec{\partial}_p \label{liouvillian}
\end{equation}
with $H(\vec{r},\vec{p}\,)$ the standard Hamiltonian of the system. The starting point of the KvN formulation of classical mechanics is the definition of a Hilbert space of complex and square integrable functions on the phase space $\psi(\vec{r},\vec{p}\,)$ whose modulus square reproduces just the standard probability density in phase space $\rho(\vec{r},\vec{p}\,)$, i.e. $|\psi(\vec{r},\vec{p}\,)|^2=\rho(\vec{r},\vec{p}\,)$. We shall call the $\psi(\vec{r},\vec{p}\,)$ as the ``KvN waves" \cite{waves1} to distinguish them from the quantum mechanical wave functions. 
The particular form of the Liouvillian of Eq. (\ref{liouvillian}), which is first order in the derivatives w.r.t. $\vec{r}$ and $\vec{p}$, implies that the Liouville equation (\ref{liouveq}) can be derived from the following equation of evolution of the KvN waves which was postulated by KvN:
\begin{equation}
\displaystyle i\frac{\partial}{\partial t} \psi \left(\vec{r},\vec{p},t \right)= 
\widehat{\cal H}\, \psi \left( \vec{r},\vec{p},t\right). \label{liouvillian2}
\end{equation}
From this equation one can derive a kernel of propagation, which can be written in terms of a path integral as follows \cite{gozzi}:
\begin{equation}
\displaystyle \langle \vec{r}, \vec{p},t| \vec{r}_{\scriptscriptstyle 0}, \vec{p}_{\scriptscriptstyle 0},t_{\scriptscriptstyle 0} \rangle 
=\int {\mathscr D}^{\prime\prime}\vec{r}\,{\mathscr D}^{\prime\prime}\vec{p}\,
{\mathscr D}\vec{\lambda}_r\,{\mathscr D}\vec{\lambda}_p\, \exp\left[i \int \textrm{d}t \;{\cal L} \right] \label{pathint}
\end{equation}
where 
\begin{equation}
{\cal L}=\vec{\lambda}_r\cdot \dot{\vec{r}}+\vec{\lambda}_p\cdot \dot{\vec{p}}-
\vec{\lambda}_r\cdot \vec{\partial}_pH+\vec{\lambda}_p\cdot\vec{\partial}_r H.
\label{lagrangiana}
\end{equation}
In Eq. (\ref{pathint}) the double prime in ${\mathscr D}^{\prime\prime}$ indicates that the integral is over paths with fixed end points and $\vec{\lambda}_r$ and $\vec{\lambda}_p$ are auxiliary variables. For further details about the variables $\lambda$ and all the formalism of the path integral approach to CM see Ref. \cite{gozzi}. Differently than in Ref. \cite{gozzi}, here we have not introduced Grassmann variables because we want to stick to the 0-form sector, where the evolution is given by the simple Liouville or KvN equation (\ref{liouveq})-(\ref{liouvillian2}). Using the path integral (\ref{pathint}) we can define the commutators between two generic functions ${O}_{\scriptscriptstyle 1}$ and ${O}_{\scriptscriptstyle 2}$ as $\displaystyle \left[{O}_{\scriptscriptstyle 1},{O}_{\scriptscriptstyle 2}\right]
\equiv\lim_{\epsilon\to 0}\langle
O_{\scriptscriptstyle 1}(t+\epsilon)O_{\scriptscriptstyle 2}(t)-O_{\scriptscriptstyle 2}
(t+\epsilon)O_{\scriptscriptstyle 1}(t)\rangle$. The only non-zero commutators among the basic variables are given by:
\begin{equation}
\displaystyle \left[r_i, \lambda_{r_j}\right] =i\delta_{ij}, \qquad  \left[p_i, \lambda_{p_j}\right]=i\delta_{ij}.  \label{conjugate}
\end{equation}
All the other commutators are zero. In particular, the commutator between $\vec{r}$ and $\vec{p}$ is zero and this confirms that the KvN is an operatorial formulation for {\it classical} and not quantum mechanics. Furthermore let us stress that Eq. (\ref{conjugate}) implies that, in the enlarged space of the KvN approach, the variables $\vec{\lambda}_r$ and $\vec{\lambda}_p$ play the role of the ``momenta" canonically conjugated to $\vec{r}$ and $\vec{p}$ respectively. One can implement Eq. (\ref{conjugate}) by choosing $\vec{r}$ and $\vec{p}$ as operators of multiplication and $\vec{\lambda}_r$ and $\vec{\lambda}_p$ as operators of derivation: $\displaystyle \vec{\lambda}_r=-i\vec{\partial}_r,  \;\vec{\lambda}_p=-i\vec{\partial}_p$. With this choice the Hamiltonian ${\cal H}$, associated to the ${\cal L}$ of Eq. (\ref{lagrangiana}), can be turned into an operator $\widehat{\cal H}$: 
\begin{displaymath}
{\cal H}\equiv \vec{\lambda}_r\cdot \vec{\partial}_pH-\vec{\lambda}_p\cdot\vec{\partial}_r H \; \longrightarrow \; \widehat{\cal H}=-i\vec{\partial}_p H(\vec{r},\vec{p}\,) \cdot \vec{\partial}_r+
i\vec{\partial}_r H(\vec{r},\vec{p}\,) \cdot \vec{\partial}_p.
\end{displaymath}
This operator $\widehat{\cal H}$ is just the Liouvillian of Eq. (\ref{liouvillian}).
This is the reason why the path integral (\ref{pathint}) can be considered as the functional counterpart of the KvN formulation of classical mechanics.

Before going on, we want to stress that the path integral (\ref{pathint}) propagates the KvN waves once they are written in the particular representation in which $\vec{r}$ and $\vec{p}$ are operators of multiplication. Actually these operators can be realized in other ways. For example, we want to mention the representation in which Eq. (\ref{conjugate}) is implemented by choosing $\vec{r}$ and $\vec{\lambda}_p$ as operators of multiplication and $\vec{\lambda}_r$ and $\vec{p}$ as $\vec{\lambda}_r=-i\vec{\partial}_r$ and $\vec{p}=i\partial/\partial\vec{\lambda}_p$ respectively, see \cite{waves1}-\cite{abr}. With this choice the KvN waves depend on $\vec{r}$ and $\vec{\lambda}_p$ and their propagation is given by the following path integral:
\begin{eqnarray}
\displaystyle \langle \vec{r}, \vec{\lambda}_p,t| \vec{r}_{\scriptscriptstyle 0}, \vec{\lambda}_{p_0},t_{\scriptscriptstyle 0} \rangle 
&\hspace{-3mm}=&\hspace{-3mm}\int {\mathscr D}^{\prime\prime}\vec{r}\,{\mathscr D}\vec{p}\,{\mathscr D}\vec{\lambda}_r{\mathscr D}^{\prime\prime}\vec{\lambda}_p\exp\left[i \int \textrm{d}t \left(\vec{\lambda}_r\cdot \dot{\vec{r}}- \vec{p}\cdot \dot{\vec{\lambda}}_p-
\vec{\lambda}_r\cdot \vec{\partial}_pH+\vec{\lambda}_p\cdot\vec{\partial}_r H \right) \right]. \nonumber\\
&\hspace{-2mm}&\hspace{-2mm} \label{pathint2}
\end{eqnarray}
If we consider a Hamiltonian of the form $\displaystyle H=p^2/2+V\left(\vec{r}\,\right)$, then Eq. (\ref{pathint2}) becomes:
\begin{displaymath}
\displaystyle \langle \vec{r}, \vec{\lambda}_p,t| \vec{r}_{\scriptscriptstyle 0}, \vec{\lambda}_{p_0},t_{\scriptscriptstyle 0} \rangle 
=\int {\mathscr D}^{\prime\prime}\vec{r}\,{\mathscr D}\vec{p}\,
{\mathscr D}\vec{\lambda}_r{\mathscr D}^{\prime\prime}\vec{\lambda}_p\, \exp\left[i \int \textrm{d}t \left(\vec{\lambda}_r\cdot \dot{\vec{r}}- \vec{p}\cdot \left(\dot{\vec{\lambda}}_p+
\vec{\lambda}_r\right)+\vec{\lambda}_p\cdot\vec{\partial}_r V \right) \right]. \end{displaymath}
The functional integral over $\vec{p}$ can be performed explicitly: it simply gives a functional Dirac delta $\delta(\dot{\vec{\lambda}}_p+\vec{\lambda}_r)$. Via this delta we can perform explicitly also the functional integral over $\vec{\lambda}_r$, simply replacing in the weight of the path integral $\vec{\lambda}_r$ with $-\dot{\vec{\lambda}}_p$. The final result is that the path integral for classical mechanics boils down to be:
\begin{equation}
\displaystyle \langle \vec{r}, \vec{\lambda}_p,t| \vec{r}_{\scriptscriptstyle 0}, \vec{\lambda}_{p_0},t_{\scriptscriptstyle 0} \rangle 
=\int {\mathscr D}^{\prime\prime}\vec{r}\,{\mathscr D}^{\prime\prime}\vec{\lambda}_p\, \exp\left[i \int \textrm{d}t \left(-\dot{\vec{\lambda}}_p\cdot \dot{\vec{r}}+\vec{\lambda}_p\cdot\vec{\partial}_r V(\vec{r}\,) \right) \right]. \label{final}
\end{equation}
This is the equation that we will use in the next sections to study the scale symmetry in the KvN formalism.

\section{Scale symmetry in classical mechanics}

Let us consider the action describing a three-dimensional non-relativistic particle of unit mass in a potential $V(\vec{r}\,)$, whose action is:
\begin{equation}
\displaystyle S=\int \textrm{d}t \, \left[ \frac{1}{2}\dot{r}^2-V(\vec{r}\,) \right]. \label{action}
\end{equation}
The scale transformation is a dilation of the time variable. In particular, if we transform time as 
$t^{\prime}=e^{-\tilde{\alpha}}t$, its infinitesimal version is: 
\begin{equation}
\displaystyle \delta t =-\tilde{\alpha} t. \label{changet}
\end{equation}
Choosing a system of units where $m$ and $\hbar$ are dimensionless, from Eq. (\ref{action}) we have that: 
\begin{equation}
[\vec{r}\,]=[t^{\scriptscriptstyle 1/2}]. \label{rrescale}
\end{equation}
As the scale transformation transforms the variables according to their physical dimension, we have from (\ref{changet}) and (\ref{rrescale}) that $\vec{r}$ rescales as:
\begin{equation}
\displaystyle \delta \vec{r}= - \frac{\tilde{\alpha}}{2}\vec{r}. \label{scaletr}
\end{equation}
Using (\ref{changet}) and (\ref{scaletr}) into (\ref{action}) one easily obtains that the only monomial potential which guarantees the scale invariance of the action (\ref{action}) is the inverse square potential: $\displaystyle V(\vec{r}\,)=-\frac{g}{2r^2}$. So the associated action is:
\begin{equation}
\displaystyle S=\frac{1}{2}\int \textrm{d}t \, \left( \dot{r}^2+\frac{g}{r^2} \right).
\label{scaleaction}
\end{equation}
Via the Noether theorem one gets that the conserved charge is the following one:
\begin{equation}
\displaystyle D=tH - \frac{1}{4}\left(\vec{r}\cdot \vec{p}+\vec{p}\cdot \vec{r}\,\right), \label{dil}
\end{equation} 
where $\vec{p}=\dot{\vec{r}}$ and $H$ is the Hamiltonian $\displaystyle H=\frac{1}{2}\left(p^2-\frac{g}{r^2}\right)$. In Eq. (\ref{dil}) we have symmetrized $\vec{r}$ and $\vec{p}$ in order to get a Hermitian charge at the quantum mechanical level. Before going on, let us notice that Eqs. (\ref{changet}) and (\ref{scaletr}) implies that the scale transformations of the momenta $\vec{p}=\dot{\vec{r}}$ are $\displaystyle \delta \vec{p}=\frac{\tilde{\alpha}}{2}\vec{p}$, i.e., the momenta $\vec{p}$ transform with the opposite sign w.r.t. $\vec{r}$. It is easy to prove that these transformations on $\vec{p}$ and $\vec{r}$ can be generated via the Poisson brackets by the dilation charge $D$ of Eq. (\ref{dil}). This means that the scale symmetry can be implemented as a {\it canonical} transformation in the standard phase space of classical mechanics.

Now let us study what happens when we implement classical mechanics via the KvN operatorial formalism. Let us indicate with ${\mathscr S}$ the weight of the path integral (\ref{final}) which, in the case of an inverse square potential $\displaystyle V(\vec{r}\,)=-\frac{g}{2r^2}$, turns out to be:
\begin{equation}
\displaystyle {\mathscr S}=\int \textrm{d}t\, \left( -\dot{\vec{\lambda}}_p\cdot \dot{\vec{r}}
+g \frac{\vec{\lambda}_p\cdot \vec{r}}{r^4} \right). \label{cpiweight}
\end{equation}
As ${\mathscr S}$ is dimensionless, we get that the dimensions of $\vec{\lambda}_p$ must satisfy the relation $[\vec{\lambda}_p]\cdot [\vec{r}\,]=[t]$, i.e.: $[\vec{\lambda}_p]=\left[t^{\scriptscriptstyle 1/2}\right]$. This means that, under a scale transformation, $\vec{\lambda}_p$ transforms exactly as the position $\vec{r}$. So the full set of transformations is given by:
\begin{equation}
\displaystyle \delta \vec{r}= - \frac{\tilde{\alpha}}{2}\vec{r}, \qquad  
\delta \vec{\lambda}_p= - \frac{\tilde{\alpha}}{2}\vec{\lambda}_p, \qquad
\delta t =-\tilde{\alpha} t. \label{transf}
\end{equation}
It is easy to check that the above transformations are a symmetry for the ${\mathscr S}$ of Eq. (\ref{cpiweight}). Implementing the Noether procedure in ${\mathscr S}$ and using the equations of motion $\dot{\vec{r}}=\vec{p}$ and $\dot{\vec{\lambda}}_p=-\vec{\lambda}_r$ we get the associated Noether charge, whose KvN operatorial version is:
\begin{equation}
\displaystyle \widehat{\cal D}=t\widehat{\cal H}+\frac{1}{2}\left(\vec{\lambda}_p\cdot \vec{p}-\vec{\lambda}_r\cdot \vec{r}\right), \label{dill}
\end{equation}
where $\displaystyle \widehat{\cal H}=\vec{\lambda}_r\cdot\vec{p}-g\frac{\vec{\lambda}_p\cdot \vec{r}}{r^4}$ is the Liouvillian of the theory. In the inverse square potential there is not only the scale invariance but a full conformal invariance \cite{alfaro}. The relative KvN operators have been built in Ref. \cite{enrico}.


\section{Quantum scale anomaly and KvN ``cancellation"}

In Ref. \cite{chir} we proved that the variables $\vec{\lambda}$ play a crucial role in producing a mechanism of ``cancellation" of the chiral anomaly in a classical field theory. In this section we want to show that something ``similar", even if not exactly the same, takes place when we study the scale transformations of Eq. (\ref{transf}).
In Ref. \cite{cambl} and \cite{gupta} it was shown that the scale symmetry associated to the inverse square potential (\ref{scaleaction}) develops an anomaly at the QM level. Basically one has to check the conservation at the quantum level of the charge $D$ of Eq. (\ref{dil}):
\begin{equation}
\frac{\textrm{d} \widehat{D}}{\textrm{d}t}= \frac{\partial \widehat{D}}{\partial t}+\frac{1}{i\hbar}[\widehat{D},\widehat{H}] = \widehat{H}+\frac{1}{i\hbar}[\widehat{D},\widehat{H}].  \label{diciotto}
\end{equation}
It was proved in Ref. \cite{cambl} that the RHS of (\ref{diciotto}) is different from zero: $\displaystyle \widehat{H} +\frac{1}{i\hbar}[\widehat{D},\widehat{H}] \neq 0$
and in particular
\begin{equation}
\displaystyle \widehat{H}+\frac{1}{i\hbar}[\widehat{D},\widehat{H}]=\widehat{A}(\vec{r}\,) \label{an}
\end{equation}
where the square brackets $[\;,\;]$ indicate the quantum commutators not to be confused with the KvN ones of Eq. (\ref{conjugate}) and $\widehat{A}(\vec{r}\,)$ in three dimensions is given by:
\begin{equation}
\displaystyle
\widehat{A}(\vec{r}\,)= \left( 1+\frac{1}{2} \vec{r}\cdot \vec{\partial}_r\right) V(\vec{r}\,)=\frac{r}{2}\,\vec{\partial}_r\cdot \left(\vec{r}\,\frac{V(\vec{r}\,)}{r}\right).
\label{anbis}
\end{equation}
For a classically scale invariant potential $\displaystyle V(\vec{r}\,)=-\frac{g}{2r^2}$ the divergence in (\ref{anbis}) is formally proportional to a three-dimensional Dirac delta and 
the final expression for the anomaly $\widehat{A}(\vec{r}\,)$ becomes \cite{cambl}:
\begin{displaymath}
\widehat{A}(\vec{r}\,) =- g\pi r\delta(\vec{r}\,).
\end{displaymath}
Note that it is the singular behavior of the inverse square potential which generates a Dirac delta in the origin.
From Eqs. (\ref{diciotto}), (\ref{an}) and the previous expression we can derive that the mean value of the dilation charge on a state $\psi$ evolves as follows:
\begin{eqnarray}
\displaystyle \frac{\textrm{d}}{\textrm{d}t} \langle \widehat{D}\rangle_{\psi} &\hspace{-2mm}= &\hspace{-2mm} \langle \widehat{A}\,
\rangle_{\psi} \nonumber \\
&\hspace{-2mm}= &\hspace{-2mm} -g\pi \int \textrm{d}\vec{r}\, \delta(\vec{r}\,) \left|r^{\scriptscriptstyle 1/2}\psi(\vec{r}\,)\right|^2. \label{anom}
\end{eqnarray}
As explained in Ref. \cite{cambl}, in the strong coupling regime there are problems in imposing the boundary condition $\psi(\vec{0}\,)=0$ which would make the RHS of (\ref{anom}) equal to zero and the anomaly disappear. To handle the RHS of (\ref{anom}) one has to introduce a regularization \cite{cambl} which leads to a breaking of the scale invariance at the QM level.

It is quite natural to use the same techniques to analyze what happens in the KvN operatorial approach to CM. Since scale symmetry holds in CM, we expect to have no anomaly factor. The operator which makes the evolution in the KvN space is the Liouvillian $\widehat{\cal H}$, so the time evolution of the charge $\widehat{\cal D}$ of Eq. (\ref{dill}) is given by the following equation, which is the analog of Eqs. (\ref{diciotto}), (\ref{an}):
\begin{equation}
\displaystyle \frac{\textrm{d}\widehat{\cal D}}{\textrm{d}t}=\widehat{\cal H}+\frac{1}{i}[\widehat{\cal D},\widehat{\cal H}]=\displaystyle \widehat{\cal A},
\label{analog}
\end{equation}
where the commutators above are the KvN ones of Eq. (\ref{conjugate}) and not the quantum ones.
What we want to do now is to calculate explicitly $\widehat{\cal A}$. Evaluating the commutator of Eq. (\ref{analog}), we get: 
\begin{eqnarray}
\displaystyle \widehat{\cal A} &\hspace{-2mm}=&\hspace{-2mm}\widehat{\cal H}+ \frac{1}{2i} \left[\left(\vec{\lambda}_p\cdot \vec{p}-
\vec{\lambda}_r\cdot \vec{r}\right),\vec{\lambda}_r\cdot \vec{p}-g\frac{\vec{\lambda}_p\cdot \vec{r}}{r^4}\,\right] \medskip \nonumber \\
\displaystyle &\hspace{-2mm}=&\hspace{-2mm} \widehat{\cal H}_{\scriptscriptstyle V}+
\frac{1}{2i}\left[\left( \vec{\lambda}_p\cdot \vec{p}-\vec{\lambda}_r\cdot \vec{r}\right),
\widehat{\cal H}_{\scriptscriptstyle V}\right] \medskip \nonumber\\
&\hspace{-2mm}=&\hspace{-2mm}
 \left(\frac{3}{2}+\frac{1}{2}\vec{r}\cdot \vec{\partial}_r\right){\widehat{\cal H}}_{\scriptscriptstyle V}, \nonumber
\end{eqnarray}
where $\displaystyle \widehat{\cal H}_{\scriptscriptstyle V}\equiv-g\frac{\vec{\lambda}_p\cdot \vec{r}}{r^4}$ is the part of the Liouvillian which comes from the inverse square potential $\displaystyle V(\vec{r}\,)=-\frac{g}{2r^2}$. 
In three dimensions $\widehat{\cal A}$ can be easily written as a divergence:
\begin{equation}
\displaystyle \widehat{\cal A}=\frac{1}{2}\vec{\partial}_r\cdot\left(\vec{r}\,\widehat{\cal H}_{\scriptscriptstyle V} \right)=
-\frac{g}{2} \vec{\partial}_r \cdot \left( \vec{r}\;
\frac{\vec{\lambda}_p\cdot \vec{r}}{r^4} \right). \label{div}
\end{equation}
Evaluating the expression (\ref{div}) one naively would get zero exactly like for the expression (\ref{anbis}). The subtleties arise in (\ref{anbis}) at $r=0$ and we could expect the same for the expression (\ref{div}). To check that we proceed to calculate the expectation value of $\widehat{\cal A}$. In evaluating the evolution of the expectation value of $\widehat{\cal D}$ on the KvN waves $\psi(\vec{r},\vec{\lambda}_p)$, we get
\begin{eqnarray}
\displaystyle \frac{\textrm{d}}{\textrm{d}t}\langle \widehat{\cal D}\rangle_{\psi}
&\hspace{-2mm}=&\hspace{-2mm} \langle \widehat{\cal A}\, \rangle_{\psi} \nonumber\\
&\hspace{-2mm}=&\hspace{-2mm} -\frac{g}{2} \int \textrm{d}\vec{\lambda}_p\left[\int \textrm{d}\vec{r}\, \vec{\partial}_r\cdot \left(\vec{r}\,\frac{\vec{\lambda}_p\cdot \vec{r}}{r^4}\right)\,|\psi|^2\right]. \label{clan}
\end{eqnarray}
We have seen before that in QM the divergence entering the RHS of Eq. (\ref{anbis}) is proportional to a Dirac delta in the origin. In CM, as there is no anomaly, we should be able to prove that the RHS of Eq. (\ref{clan}) is zero. 
First of all, it is straightforward to show that the divergence of Eq. (\ref{div}) is zero for all the points with $\vec{r}\neq \vec{0}$. Consequently, we can reduce the integral over $\vec{r}$ of Eq. (\ref{clan}) to an integral over a sphere $S_{\varepsilon}$ of radius $\varepsilon$. This radius can be taken arbitrary small, in such a way that the variations of $|\psi|^2$ over the sphere $S_{\varepsilon}$ become negligible. By using the divergence theorem, the volume integral can be reduced to an integral over the surface $\sigma$ of the sphere $S_{\varepsilon}$:
\begin{displaymath}
\displaystyle 
\int_{S_{\varepsilon}} \textrm{d}\vec{r} \; \vec{\partial}_r\cdot \left(\vec{r}\;\frac{\vec{\lambda}_p\cdot \vec{r}}{r^4} \right) =
\int_{\sigma} \textrm{d}\vec{n} \cdot \left(\vec{r}\;\frac{\vec{\lambda}_p\cdot \vec{r}}{r^4} \right).
\end{displaymath}
If we evaluate the last integral above in spherical coordinates then the ray $\varepsilon$ disappears from the integral and we get:
\begin{displaymath}
\displaystyle 
\int \, \sin \theta \,\textrm{d}\theta \, \textrm{d}\varphi \,  \left(\lambda_{p_x} 
\cos \varphi\sin\theta+\lambda_{p_y} \sin \varphi\sin \theta +\lambda_{p_z} \cos \theta\right)=0,
\end{displaymath}
where we have used the periodicity of the trigonometric functions over the spherical coordinates $\theta\in[0,\pi)$ and $\varphi\in[0,2\pi)$. So, while in the quantum case the divergence $\displaystyle \vec{\partial}_r \cdot \left( \vec{r}\frac{1}{r^3} \right)$ of Eq. (\ref{anbis}) is formally proportional to a Dirac delta, in CM the coupling between $\vec{r}$ and $\vec{\lambda}_p$ makes the expression $\displaystyle \vec{\partial}_r\cdot \left(\vec{r}\, \frac{\vec{\lambda}_p\cdot \vec{r}}{r^4}\right)$ equal to zero.
Inserting this result into Eq. (\ref{clan}), we can conclude that both $\langle \widehat{\cal A}\, \rangle_{\psi}$ and $\displaystyle \frac{\textrm{d}}{\textrm{d}t}\langle \widehat{\cal D}\rangle_{\psi}$ are identically zero, i.e. the mean value of the KvN dilation charge is conserved in time. This concludes the proof that in the case of the inverse square potential no scale anomaly appears in the 
KvN formulation of CM. 

Another method to prove the absence of anomalies in the KvN formalism is based on the study of the self-adjointness of the Hamiltonian. The method was started for QM by Esteve in Ref. \cite{esteve2} and continued in Ref. \cite{esteve}. We will skip the details here and refer the reader to those papers. In our case we have to study the self-adjointness of the $\widehat{\cal H}$ appearing in Eq. (\ref{dill}). This study is performed by calculating the deficiency indices $d_{\pm}$. They are defined \cite{reed} as the dimension of the Hilbert space given by the solutions of the equation
\begin{displaymath}
\displaystyle \widehat{\cal H}\psi(\vec{r},\vec{p}\,)=\pm i\psi(\vec{r},\vec{p}\,).
\end{displaymath}
A solution of this equation is:
\begin{equation}
\displaystyle \psi(\vec{r},\vec{p}\,)=\widetilde{D}\left(\frac{p^2}{2}-\frac{g}{2r^2}\right) \exp \Biggl[\mp \frac{1}{2}\left(\vec{p}\cdot \vec{r}\,\right)\left(\frac{p^2}{2}-\frac{g}{2r^2}\right)^{-1}\Biggr]. \label{ultima}
\end{equation}
We have strong indications that this is the only solution. We immediately notice that (\ref{ultima}) are not square integrable functions in phase space, which implies that $d_{\pm}$ are zero. This proves that $\widehat{\cal H}$ is self-adjoint and, as a consequence, that there are no anomalies as explained by Esteve \cite{esteve2}. 

Before concluding this paper, we want to stress again that it was the presence of the auxiliary variables $\lambda$ in the $\widehat{\cal A}$ of Eq. (\ref{div}) which made the RHS of (\ref{clan}) zero. So, like for the chiral anomaly analyzed in Ref. \cite{chir}, the $\lambda$ seem to play some role in ``cancelling" quantum effects like the anomalies. This is not so surprising. In fact it was proved in Ref. \cite{abr} that the quantum fluctuations are ``somehow" killed by the extension of the usual Hilbert space of QM to an enlarged space which contains the $\lambda$.

\section*{Acknowledgments}
We would like to thank M. Reuter and K.S. Gupta for useful discussions. This research has been supported by grants from INFN, MIUR and the University of Trieste.

\end{document}